\documentclass[aps,prl,twocolumn,superscriptaddress,groupedaddress,preprintnumbers]{revtex4-2}  

\usepackage{amsmath}
\usepackage{amsfonts}
\usepackage{amssymb}
\usepackage{graphicx}
\usepackage{dcolumn}
\usepackage{bm}
\usepackage[normalem]{ulem}
\usepackage{xcolor}

\definecolor{darkspringgreen}{rgb}{0.09, 0.45, 0.27}
\definecolor{seagreen}{rgb}{0.18, 0.55, 0.34}
\definecolor{cadmiumgreen}{rgb}{0.0, 0.42, 0.24}

\usepackage[colorlinks=true, allcolors=seagreen]{hyperref}

\usepackage[english]{babel}

\def \aap  {A\&A}

\def \apj  {ApJ}

\def \apjl  {ApJL}

\def \araa  {ARA\&A}

\def \mnras {MNRAS}

\def \aapr {A\&A Rev.}

\def\beq{\begin{equation}}
\def\eeq{\end{equation}}
\def\beqa{\begin{eqnarray}}
\def\eeqa{\end{eqnarray}}

\begin{document}

\title{The Sun as a target for axion dark matter detection}

\author{Elisa Todarello,} 
\email{elisamaria.todarello@unito.it}
\affiliation{Dipartimento di Fisica, Universit\`a di Torino, Via P. Giuria 1, 10125 Torino, Italy}
\affiliation{Istituto Nazionale di Fisica Nucleare, Sezione di Torino, Via P. Giuria 1, 10125 Torino, Italy}
\author{Marco Regis} 
\email{marco.regis@unito.it}
\affiliation{Dipartimento di Fisica, Universit\`a di Torino, Via P. Giuria 1, 10125 Torino, Italy}
\affiliation{Istituto Nazionale di Fisica Nucleare, Sezione di Torino, Via P. Giuria 1, 10125 Torino, Italy}
\author{Marco Taoso} 
\email{taoso@to.infn.it}
\affiliation{Istituto Nazionale di Fisica Nucleare, Sezione di Torino, Via P. Giuria 1, 10125 Torino, Italy}
\author{Maurizio Giannotti} 
\email{mgiannotti@unizar.es}
\affiliation{Centro de Astropart{\'i}culas y F{\'i}sica de Altas Energ{\'i}as (CAPA), Universidad de Zaragoza, Zaragoza, 50009, Spain}\
\email{jvogel@unizar.es}
\author{Jaime Ruz} 
\email{jruz@unizar.es}
\affiliation{Centro de Astropart{\'i}culas y F{\'i}sica de Altas Energ{\'i}as (CAPA), Universidad de Zaragoza, Zaragoza, 50009, Spain}
\author{Julia K. Vogel} 
\email{jvogel@unizar.es}
\affiliation{Centro de Astropart{\'i}culas y F{\'i}sica de Altas Energ{\'i}as (CAPA), Universidad de Zaragoza, Zaragoza, 50009, Spain}

\begin{abstract}
    The exploration of the parameter space of axion and axion-like particle dark matter is a major aim of the future program of astroparticle physics investigations. 
    In this context, we present a possible strategy that focuses on detecting radio emissions arising from the conversion of dark matter axions in the Sun's magnetic field, including conversion in sunspots. We demonstrate that near-future low-frequency radio telescopes, such as the SKA Low, may access regions of unexplored parameter space for masses $m_a\lesssim 10^{-6}$ eV. 
\end{abstract}
\maketitle

QCD axions are hypothetical particles introduced to solve the strong CP problem, namely the non-observation of CP violation from QCD interactions~\cite{Peccei:1977hh,Peccei:1977ur, Weinberg:1977ma, Wilczek:1977pj}, see e.g., refs.~\cite{Kim:2008hd, DiLuzio:2020wdo} for recent reviews.
They are pseudo-scalar Nambu-Goldstone bosons associated with the spontaneous breaking of the so-called Peccei-Quinn symmetry. Additionally, QCD axions are also a viable dark matter (DM) candidate, possibly accounting for the entirety of cold DM in the Universe.
They can be produced through different thermal and non-thermal mechanisms in the early Universe, before or after inflation~\cite{Preskill:1982cy,Abbott:1982af,Dine:1982ah}.
The allowed range of the Peccei-Quinn scale, $f_a$, to obtain the observed DM relic density is $10^{10} - 10^{12}$ GeV, corresponding to masses $m_a$ in the rage $10^{-6} - 10^{-3}$~eV. 
The QCD DM axion stands as a key focus in the ongoing and future program of astroparticle physics investigations. Numerous experimental searches, both in the laboratory and in the sky have been designed to look for this particle~\cite{Chou:2022luk,Adams:2022pbo,Antel:2023hkf}.

Furthermore, several beyond-the-standard-model theories, such as string theory models~\cite{Arvanitaki:2009fg}, predict in many cases 
 one or more axion-like particles (ALPs), with properties very similar to those of axions but not necessarily related to the strong CP-problem~\cite{Jaeckel:2010ni}.
ALPs can have masses as low as $10^{-22}$ eV and are typically very weakly coupled to the SM. 
Moreover, they are also good DM candidates in some portions of the parameter space~\cite{Arias:2012az}. 

Experimental searches for both the QCD axion and ALPs often rely on their coupling to photons~\cite{Sikivie:1983ip}.
In both cases, the coupling is described by the effective Lagrangian 
term $\mathcal{L} \propto g_{a\gamma} a F_{\mu \nu}\tilde{F}^{\mu \nu}$.

One of the most promising ALP~\footnote{From now on, we will use the terms axion and ALP interchangeably.} searches with astrophysical probes involve the radio signal associated with ALP-photon conversion in magnetized astrophysical plasma. This signature is based on the above-mentioned ALP-photon coupling, and it would appear as a nearly monochromatic spectral line.
In order to have a significant probability of conversion, large magnetic fields and dense plasma are required so to resonantly amplify the process.
In the past few years, the search has been focused especially on the magnetosphere of 
neutron stars~\cite{Pshirkov:2007st,Hook:2018iia,Safdi:2018oeu,Huang:2018lxq,Battye:2019aco,Leroy:2019ghm,Foster:2020pgt,Prabhu:2020yif,Witte:2021arp,Battye:2021xvt,Battye:2021yue,Wang:2021hfb,Foster:2022fxn,Zhou:2022yxp,McDonald:2023shx,McDonald:2023ohd,Tjemsland:2023vvc,Buckley_2021}. 

In this paper, we instead highlight that the Sun and its sunspots are promising targets.
Sunspots are magnetic structures located within active regions of the Sun.
They appear darker on the solar surface than the normal solar photosphere (quiet Sun) and are composed of an
inner, darker part, called the umbra, and an outer, less dark part, called the penumbra.
It is well known that the magnetic field can be very intense above the photosphere of sunspots, reaching a few thousand Gauss in the umbra~\cite{2003A&ARv..11..153S}. 
The magnetic field decreases with height in the solar atmosphere, but it could still be rather strong in the chromosphere and corona. 
For example, Ref.~\cite{Anfinogentov:2019} reported a magnetic field of about 4000 G at the base of the corona above a sunspot, with 1000 G at a height of $10^4$ km. Sunspots might be environments hosting copious axion-photon conversions in their magnetic field.

The idea of considering the Sun was proposed in \cite{An:2020jmf} for a similar process, i.e. investigating dark photon-photon mixing. A search for the associated monochromatic radio signal in the solar observation data of the LOFAR telescope was then conducted in \cite{An:2023}, which also discusses the application to the axion-photon conversion.

Besides non-relativistic DM axions, one could also consider the conversion of relativistic axions thermally produced in the Sun's interior and then converted into photons in the Sun's atmosphere~\cite{Carlson:1995xf}. 
In this case, they would lead to X-ray signatures, which we plan to investigate in a forthcoming analysis~\cite{xray}.

This paper aims to provide a simple and general description of the radio emission arising from the conversion of ambient DM axions in the Sun's magnetic field. 
The goal is to present robust and broad arguments with computations that can be easily reproduced by the reader.
These arguments should provide a framework that can later be applied to specific examples in dedicated studies.  
As we shall see, the analysis of the radio signal from axion conversion in the solar magnetic field may provide an opportunity to access regions of unexplored parameter space with the next generation of radio telescopes. 

For a non-relativistic axion,  a weakly magnetized plasma, and assuming the magnetic field to be static, the conversion probability can be approximated as~\cite{McDonald:2023ohd}
\beq
P_{a\rightarrow \gamma}\simeq\frac{\pi}{2}\,\frac{g_{a\gamma}^2\,B_\perp^2}{v_a\,\omega^\prime_{q|res}}\;,
\label{eq:prob}
\eeq
where $\omega_q$ is the plasma frequency and $\omega^\prime_{q|res}$ is its gradient $d\omega_q/dr$ evaluated at the resonance location, i.e., at the radius $r_c$ where $\omega_q\simeq m_a$. In eq.~(\ref{eq:prob}) and in the following, we assume spherical symmetry and, for simplicity, radial emission.
The radius $r_c$ at which the conversion occurs, for different ALP masses, can be estimated from the resonance condition and the relation $\omega_q(r)=1.17\,\mu{\rm eV}\,\sqrt{n_e(r)/(10^9\,{\rm cm^{-3}})}$, where $n_e$ is the electron plasma density.

For a uniform conversion over a surface $\Delta A$, and assuming the ALP distribution to be isotropic, the ALP-induced photons flux per unit frequency on Earth can be estimated as
\beqa
S&=&\int\,\frac{d\Omega}{4\pi\,\Delta\nu}\,\rho_a\,v_a\,P_{a\rightarrow \gamma}\,e^{-\tau}\nonumber\\
&\simeq&\frac{\Delta A}{4\pi\,\,\Delta\nu\,d^2}\,\rho_a\,v_a\,P_{a\rightarrow \gamma}\,e^{-\tau}\;,
\label{eq:flux}
\eeqa
 where $d$ is the distance of the conversion surface from us, $\rho_a$ is the DM density at the conversion surface, $\tau$ is the photon optical depth, and $\Delta\nu$ is the bandwidth of the signal.

To gain some intuition about the expected signal, let us consider a sunspot, i.e., a structure with strong magnetic fields appearing on the solar surface. Considering a 
radius of the sunspot $\ell_s$, the conversion area is computed as $\Delta A=\pi\,\ell_s^2.$
For an analytical estimate, we can approximate the plasma profile as a power-law $\omega_p\propto h^\alpha,$ with $\alpha\simeq0.5$ in the range of conversion radii of interest. Then we have $\omega^\prime_{q|res}=\alpha\,\omega_p/h_c,$ where $h_c$ is the distance of the conversion surface from the photosphere.  
Plugging in typical values for a large sunspot~\cite{2003A&ARv..11..153S}, we find

\beqa
S&=&\frac{\Delta A}{8\,\,\Delta\nu\,d^2}\,\rho_a\,\frac{g_{a\gamma}^2\,B^2}{\omega^\prime_{q|res}}\,e^{-\tau}=0.7\,{\rm mJy}\,\left(\frac{10^{-6}}{\Delta \nu/\nu}\right)\nonumber\\
&\times&\left(\frac{\ell_s}{4\times 10^4{\rm km}}\right)^2\,\left(\frac{\rho_a}{1.0\,{\rm GeV/cm^3}}\right)
\left(\frac{g_{a\gamma}}{10^{-12}{\rm GeV^{-1}}}\right)^2\nonumber \\
&\times&\left(\frac{B_\perp}{10\,{\rm G}}\right)^2\,
\left(\frac{\mu{\rm eV}}{m_a}\right)^2\,\left(\frac{0.5}{\alpha}\right)\,\left(\frac{h_c}{3\times 10^3{\rm km}}\right)\,e^{-\tau}\;.
\label{eq:forec}
\eeqa

\begin{figure}
    \centering
    \includegraphics[width=0.48\textwidth]{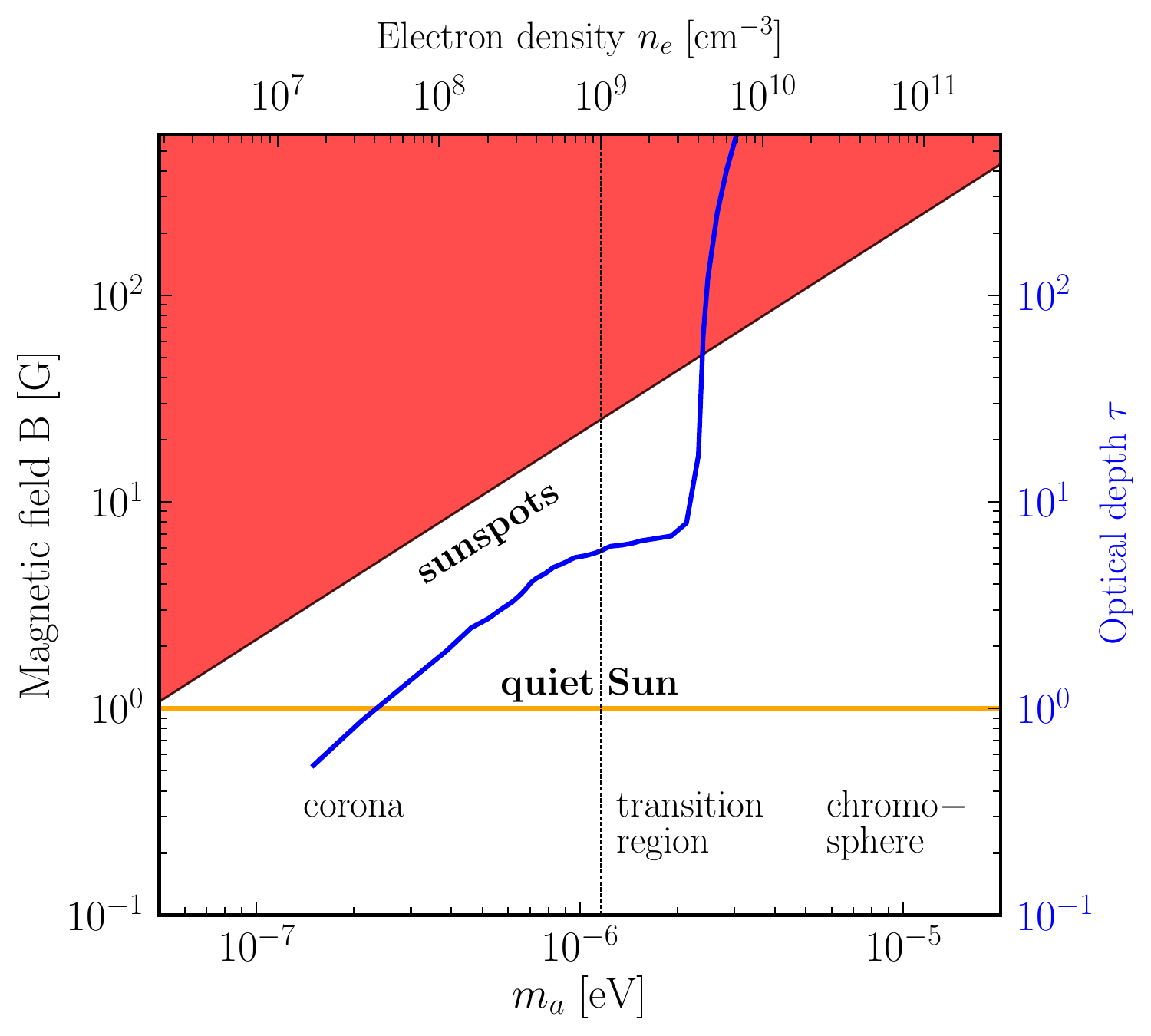}
    \caption{In the red region, the electron cyclotron resonance causes a strong absorption of the photons produced by ALPs conversion during their propagation in the solar atmosphere. Assuming that the magnetic field decreases monotonically with distance from the photosphere, this effect does not occur when the magnetic field at the ALP conversion location lies in the white region. The optical depth associated with thermal bremsstrahlung, the other main absorption mechanism, is shown in blue (right-axis). The top axis shows the electron density at which the 
    plasma frequency matches the ALP mass (bottom axis), and the resonant conversion can occur. Vertical lines mark different regions of the solar atmosphere. The orange horizontal line marks the typical value of 1~G of the magnetic field in the corona for the quiet Sun. 
    }
    \label{fig:phys}
\end{figure}
In the previous equation, we have considered a local DM density $\rho_{\infty}=0.3\,\rm{ GeV cm}^{-3}$ far away from the Sun and accounted for the gravitational focusing of the Sun following~\cite{Alenazi:2006wu}. This leads to a DM density at the solar surface of $\rho_a\simeq 1\,\rm{GeV cm}^{-3}.$
Moreover, taking a local DM velocity dispersion $v\simeq10^{-3}$ leads to a intrinsic bandwidth $\Delta\nu\simeq10^{-6}.$ 
In the following, in our numerical analysis performed to derive the sensitivity forecasts, the plasma frequency and its gradient are computed from the solar electron density distribution of Ref.~\cite{2005A&A...433..365S}. 
We find that adopting a different model, e.g.~\cite{2008GeofI..47..197D}, leads to a similar sensitivity on the axion signal.

Let us now analyze the absorption term. 
There are two main contributions to the opacity in the solar atmosphere at these wavelengths: 
the gyro-resonance absorption and the thermal bremsstrahlung. 
The first effect occurs when the resonant condition $m_a\simeq n\,\Omega_B(r)$ is met, 
where $\Omega_B=e\,B/m_e\simeq 0.012\,\mu{\rm eV}\,B/{\rm G}$ is the electron cyclotron frequency and the integer $n$ corresponds to the different energy levels.
To compute the location where the cyclotron resonance is realized, we assume that the magnetic field decays with height following a power-law behavior after the ALP conversion point. Then, we extract the electron number density and the temperature of the plasma at the cyclotron resonance point. Finally, with these ingredients and following Ref.~\cite{1985ARA&A..23..169D}, we obtain the cyclotron optical depth.
We find that for any reasonable choice of the magnetic field profile in the solar atmosphere, the optical depth $\tau_g$ is very large for the first three transitions $n=1,2,3,$ while $\tau_g\lesssim1$ for $n=4$ and negligible for higher levels. 
Therefore, assuming a monotonically decreasing magnetic field and plasma density, we conservatively assume that photons from axion conversion can reach the Earth only if the conversion happens after the surface of gyro-resonance absorption associated with the $n=4$ level.
This condition translates into an upper limit of the magnetic field at the ALP conversion point, which is represented in fig.~\ref{fig:phys}, with the \emph{no-signal region} shown in red. 

Concerning the thermal bremsstrahlung, the absorption coefficient reads~\cite{Rybickibook}:
\beq
\alpha_b\simeq 0.018\, \frac{n_e}{10^6\,{\rm cm^{-3}}}\frac{Z^2_{eff}n_i}{10^6\,{\rm cm^{-3}}}\left(\frac{T}{{\rm K}}\right)^{-3/2} \left(\frac{\nu}{{\rm MHz}}\right)^{-2}G_F\;,
\eeq
where $n_i$ is the number density of ions with effective charge $Z_{eff}$, $T$ is the temperature of the plasma, all taken from \cite{2005A&A...433..365S}, and $G_F$ is the Gaunt factor, for which we referred to \cite{Chluba:2020}. In the corona, the plasma is essentially fully ionized, and $Z_{eff}
\approx 1$, while this is not the case in the chromosphere and transition region.
The bremsstrahlung optical depth $\tau_b=\int_{h_c}\,d\ell\,\alpha_b$ is shown as a blue line in fig.~\ref{fig:phys}. 
In the chromosphere, we have $\tau_b\gg 1$ therefore the ALP signal is completely absorbed.
On the other hand, in the corona $\tau_b\lesssim \mathcal{O}(1)$.

A crucial ingredient is the magnetic field in the corona, which determines the conversion probability as described in eq.~(\ref{eq:prob}).
The magnetic field of the quiet Sun has a strength between $1-4$ G, on average~\cite{Yang:2020Sci}. 
We will take $B=1$~G when estimating the ALP signal from across the whole Sun.
As mentioned before, the magnetic field of sunspots can be very intense, even above the photosphere. Sunspot observations accumulated over the years show great variability of their properties, with sizes ranging from $10$ to $10^5$ km, and magnetic fields from few G up to few thousand G~\cite{2003A&ARv..11..153S}.

Here, in the spirit of providing a general treatment, we will assume sunspots to populate the white triangle in fig.~\ref{fig:phys}, delimited by the red region and orange line. 
Namely, we assume that there exist sunspots with associated magnetic fields in the corona and transition region, 
where the resonance ALP conversion occurs, larger than the magnetic field of the quiet Sun, and up to the maximal value allowed by the cyclotron absorption, see fig.~\ref{fig:phys}.
For the sake of concreteness, in the following, we will consider two cases: one where the magnetic field maximizes the signal, i.e. lies at the border between the red and white regions in fig.~\ref{fig:phys}, and one where we simply assume $B=4$~G at the ALP conversion radius above the sunspot.

To examine observational prospects, we derive forecasts for the first phase of the SKA observatory, which is expected to be completed in its final configuration in 2029. We follow~\cite{SKA:2019}, in particular their Table 6 provides the line sensitivity for sources with an angular size within a certain range. For larger extensions, we assume a linear degradation of the sensitivity with the angular size of the source, as suggested by their Table 4 for our range of interest. 

The angular radius of a sunspot is typically smaller than a few arcmin, e.g. $\ell_s=4\times 10^4$ km, which we will use for our estimates, corresponds to $0.9^\prime$. 
On the other hand, radio photons are affected by scattering processes during their propagation in the solar plasma, which leads to an angular smearing of the signal.
We estimate the broadening of a point-like source as described in \cite{Kontar:2019}. 
In the case of the entire Sun, this broadening is instead marginal. The same conclusion has also been reported in \cite{Sharma:2020}, finding that the observed radio size of the Sun at 100-240 MHz is only 25–30\% larger in area than the intrinsic one.

Finally, the line sensitivity reported in~\cite{SKA:2019} refers to observations of regions of the sky without strong radio emissions. Therefore, to include the large radio background produced by the Sun, we correct the quoted sensitivity adding the flux density of the Sun to the system-equivalent flux density of an SKA station. Concretely, we follow the estimates provided in~\cite{Macario:2022}. 
From Fig. 8 in~\cite{SKA:2019}, we can read the sensitivity of the SKA-Low array in terms of $A_{eff}/T_{\rm sys}$ and convert it into the System Equivalent Flux Density $SEFD=2k_B T_{\rm sys}/A_{eff}$. Since the array is composed by 512 stations, the $SEFD_s$ of an SKA station (which sees the Sun as a point-like source) is about 512 times the $SEFD$ of the SKA array. 
To derive the sensitivity, the measured $SEFD_S$ including any emission along the direction of observation has to be added. In this case the Sun gives the dominant contribution, and the degradation of the sensitivity with respect to a region with low radio emission is given by the factor $(SEFD_s+SEFD_S)/SEFD_s$.

\begin{figure}
    \centering
    \includegraphics[width=0.48\textwidth]{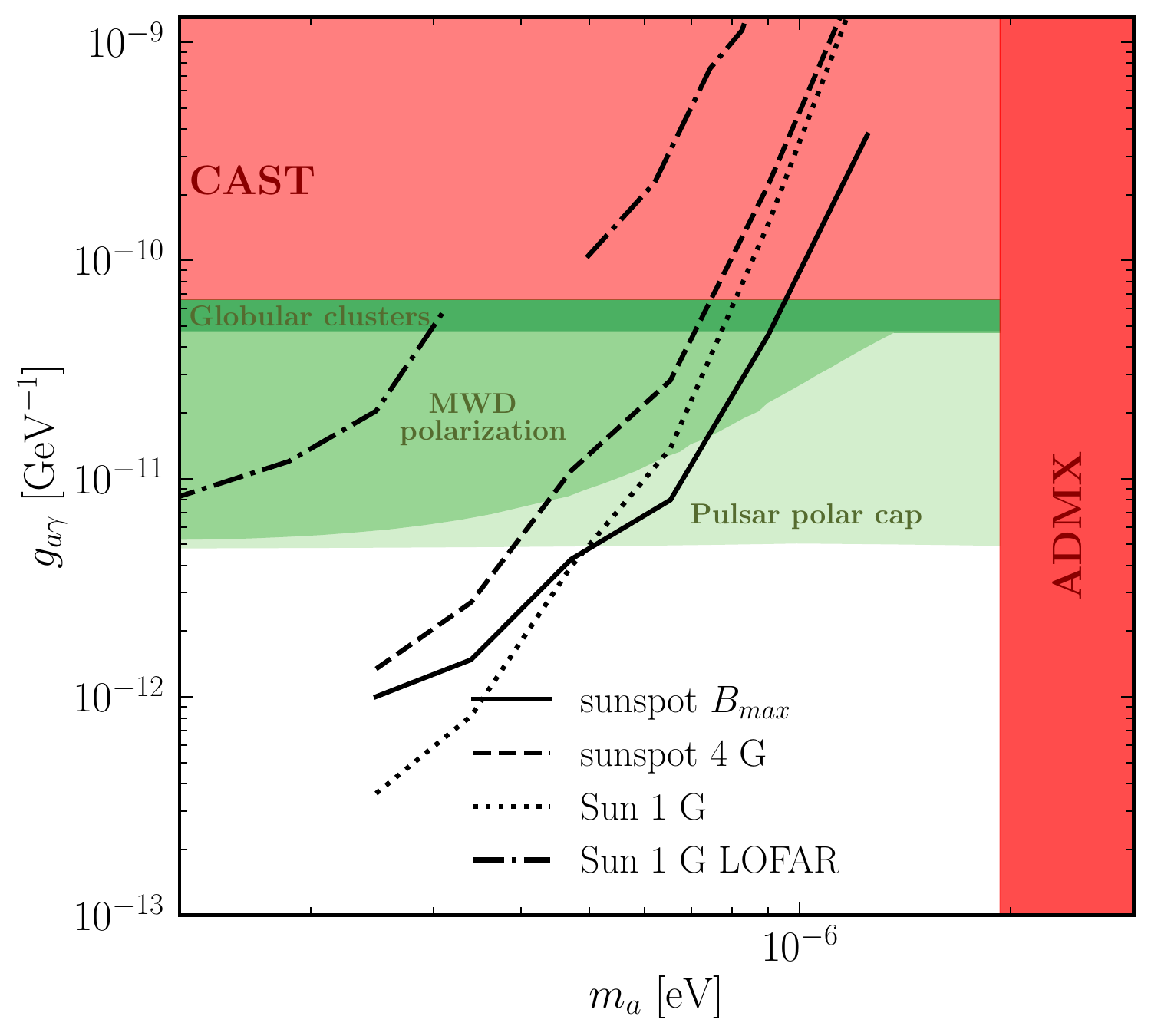}
    \caption{Projected sensitivity on the ALP-photon coupling $g_{a\gamma}$ versus ALP mass $m_a$, assuming 100 hours of observations with SKA-1 Low, of a sunspot having the maximum magnetic field allowed by the gyro-resonance absorption condition at the conversion location (see fig.~\ref{fig:phys}) (solid), a sunspot with a magnetic field of 4~G at the conversion location (dashed), and the whole Sun (dotted). The radius of the conversion surface above a sunspot is taken to be $\ell_s=4\times 10^4$ km. The dashed-dotted line shows the projected sensitivity for the whole Sun, in the case of a 8hr-observation with LoFAR~\cite{LOFAR:2013}. In red, we show the laboratory bounds from ADMX~\cite{Asztalos_2010} and CAST~\cite{CAST2017}, while in green we show astrophysical bounds: the R2-parameter in globular cluster~\cite{Dolan_2022}, and the most conservative limits from polarization of magnetic dwarf~\cite{Dessert_2022} and ALP production in pulsar polar caps~\cite{noordhuis2023novel}. Other relevant constraints in this mass range (not shown in the figure for clarity) come from the R-parameter in globular clusters~\cite{Ayala_2014}, X-ray observations of magnetic white dwarfs~\cite{Dessert_2022X} and from $\gamma$-ray observation of the blazar Markarian 421 (Mrk 421)~\cite{Li_2021, Li_2022}. 
   }
    \label{fig:res}
\end{figure}

Following the procedure explained above, we obtain the results presented in fig.~\ref{fig:res}. The projected sensitivity is at 95\% C.L., namely, they correspond to couplings providing a conversion signal equal to twice the SKA sensitivity.
For ALP masses in the range $10^{-6}-10^{-7}$ eV, the proposed search strategy can improve existing laboratory bounds, and it is competitive with existing astrophysical constraints. We find that the forecasted sensitivities for observations of the entire Sun and a sunspot with a large magnetic field are comparable. A technical advantage in the latter case is offered by the small angular size of the emission region, in contrast to the case of the entire Sun, where the large angular extension requires care in the data manipulation.
On the other hand, the properties of a sunspot needed to compute the prediction of the expected flux might not be straightforward to be derived from the observations.

The curves in fig.~\ref{fig:res} are computed assuming an observing time of $\Delta t=100$ hr. Considering that solar physics is one of the key scientific goals of the SKAO~\cite{Skasun:2019}, hundreds of hours of observations of the Sun are foreseen. In the case of sunspots, the $100$ hr value is guaranteed if intended as a combination of different sunspots, whilst it is probably difficult to achieve for a single one. Note that, in any case, the sensitivity to the ALP-photon coupling scales very moderately with observing time, as $\Delta t^{1/4}$. 

In fig.~\ref{fig:res}, we assume a bandwidth of $\Delta \nu/\nu=10^{-6}$, matching the one induced by the dispersion of the DM velocity. This is achievable by the SKA Low. On the other hand, this value might be somewhat optimistic since it neglects possible smearing effects~\footnote{We postpone a full ray-tracing computation to simulate photon propagation in the Sun to future work. Here, we accounted for absorption and spatial broadening as described in the text. The bandwidth smearing is instead neglected for simplicity.}. Note, however, that, again, the sensitivities scale as $(\Delta \nu/\nu)^{1/4}$, so even significantly larger bandwidths can still lead to interesting prospects.  

Let us also mention that the proposed SKAO observation requires a high level of dynamic range, being $4\times 10^5$ at the smallest frequency/mass up to $3\times 10^6$ at the highest frequency/mass.
A dynamic range of the order of $10^6$ is challenging, and it certainly requires significant care and effort in the data calibration and analysis. However, it is in the range of values reported in the literature~\cite{Smirnov:2011}, and such a spectral dynamic range is of the same order required for observations of the 21 cm line from the epoch of reionization, which have made steady progress towards achieving it~\cite{Barry:2021szi,HERA:2021noe}.

For completeness, we report the 
sensitivity that can be obtained with current instruments.
We focus on the LoFAR telescope~\cite{LOFAR:2013} and derive prospects for a typical observation, i.e., the 24-station array, 8 hours of integration time and 100 kHz/channel of frequency resolution (an analysis with available data and weaker sensitivity is described in~\cite{An:2023}). The forecast computation proceeds as discussed above for the SKAO. We can see from Fig.~\ref{fig:res} that the projected sensitivity falls in a strongly constrained region of the parameter space. The higher sensitivity of the SKAO and/or a dedicated campaign seem to be required to test viable models.

Let us notice that other bounds derived from radio observations, e.g., the one from \cite{noordhuis2023novel}, can also be improved by SKAO observations, possibly surpassing the projected sensitivity described here. 
However, they rely on different assumptions, and the Sun can be seen as a complementary probe.

To conclude, in this work, we highlighted that the Sun can be a promising target for looking for radio waves from ALP conversion. A discussion in this direction was recently reported also in \cite{An:2023}. 
Here we presented a simple and general derivation of the expected signal, with forecasts for the SKA, and added the case of sunspots. The Sun can be thus considered as complementary/alternative to the case of neutron stars, which are the target that has been extensively investigated in connection to the ALP conversion~\cite{Pshirkov:2007st,Hook:2018iia,Safdi:2018oeu,Huang:2018lxq,Battye:2019aco,Leroy:2019ghm,Foster:2020pgt,Prabhu:2020yif,Witte:2021arp,Battye:2021xvt,Battye:2021yue,Wang:2021hfb,Foster:2022fxn,Zhou:2022yxp,McDonald:2023shx,McDonald:2023ohd,Tjemsland:2023vvc,Buckley_2021}. 
As a simple comparison let us consider the case of a nearby isolated neutron star, keeping in mind however that neutron stars could also be located in environments hosting DM densities much larger than the local one, such as the galactic center. Focusing on the main ingredients determining the radio flux, we have
\beqa
S&\propto&\left(\frac{h_{c,s}\,\ell_s^2}{5\times 10^{12}\,{\rm km^3}}\right)\left(\frac{B_s}{5\,{\rm G}}\right)^2\left(\frac{1.5\times 10^{8}\,{\rm km}}{d_s}\right)^2\nonumber\\
&\simeq&\left(\frac{r_{c,{NS}}}{200\,{\rm km}}\right)^3\left(\frac{B_{NS}}{10^{12}\,{\rm G}}\right)^2\left(\frac{1\,{\rm kpc}}{d_{NS}}\right)^2\;,
\eeqa
where $s$ denotes a large sunspot and $NS$ a neutron star. 
Therefore, naively, we see that the expected signals are roughly comparable in order of magnitude.  
However, there are important differences between the two systems, 
in terms of systematic uncertainties and values of the relevant parameters. 
Hence, the two approaches present complementary ways to explore the parameter space corresponding to ALP dark matter. 
Since the properties of the Sun are much better known and certainly more easily accessible to future measurements, what is presented here could be a more robust and accessible method to probe the ALP parameter space in the sub $\mu$eV region.
Unfortunately, absorption effects limit the range of masses that can be probed to $m_a\lesssim 10^{-6}$ eV, as discussed in the text.
This also implies that low-frequency radio observations will be the key to test the hypothesis discussed here.

\section*{Acknowledgements}
We would like to thank Gianni Bernardi for his help in determining the SKA sensitivity, and Sam Witte for useful comments.

MT acknowledges the research grant ``Addressing systematic uncertainties in searches for dark matter No. 2022F2843L'' funded by MIUR. ET, MR, and MT acknowledge the project ``Theoretical Astroparticle Physics (TAsP)'' funded by Istituto Nazionale di Fisica Nucleare (INFN).
MR and ET acknowledge support from the research grant `From Darklight to DM: understanding the galaxy/matter connection to measure the Universe' No.\ 20179P3PKJ funded by \textsc{miur}, and from the ``Grant for Internationalization" of the University of Torino.
ET thanks the University of Zaragoza for its hospitality during the initial phases of this work.
This article/publication is based upon work from COST Action COSMIC WISPers CA21106, supported by COST (European Cooperation in Science and Technology)”. 

\bibliography{biblio}
\end{document}